\documentclass{PoS_arXiv}
\pdfoutput=1

\usepackage{graphicx} 
\graphicspath{{figs/}} 

\DeclareGraphicsExtensions{.pdf}  

\usepackage{bbm}                  
\usepackage{multirow}             
\usepackage{feynmp}
\usepackage{amsmath}              
\usepackage{amssymb}              
\usepackage{amsthm}               
\usepackage{rotating}             
\usepackage{xcolor}               
\usepackage{cite}
\usepackage{url}

\newcommand{\lsim}
{\;\raisebox{-.3em}{$\stackrel{\displaystyle <}{\sim}$}\;}

\newcommand{\id}{{\rm 1\kern-.12em
\rule{0.3pt}{1.5ex}\raisebox{0.0ex}{\rule{0.1em}{0.3pt}}}}

\newcommand{\SLASH}[2]{\makebox[#2ex][l]{$#1$}/}
\newcommand{\pslash}{\SLASH{p}{.2}}

\newcommand\tb{\tan\beta}

\newcommand{\sinb}{\sin \beta\,}
\newcommand{\cosb}{\cos \beta\,}

\newcommand{\Sbe}{\sin \beta}
\newcommand{\Cb}{\cos \beta}

\newcommand{\OP}{\omega_+}
\newcommand{\OM}{\omega_-}
\newcommand{\OMP}{\omega_{-/+}}

\newcommand\LP{\left(}
\newcommand\RP{\right)}
\newcommand\LB{\left[}
\newcommand\RB{\right]}
\newcommand\LV{\left\{}
\newcommand\RV{\right\}}

\newcommand\ReDiag{\mathop{%
  \raise .5pt\hbox{[}%
  \widetilde{\mathrm{Re}}%
  \raise .5pt\hbox{]}}}
\newcommand\ReOffDiag{\mathop{%
  \raise .5pt\hbox{$\llbracket$}%
  \widetilde{\mathrm{Re}}%
  \raise .5pt\hbox{$\rrbracket$}}}
\newcommand\SE[1]{\Sigma_{#1}}
\newcommand\rSE[1]{\hat{\Sigma}_{#1}}

\newcommand\OS{\mathrm{OS}}
\newcommand\Os{\mathrm{os}}
\newcommand\DRbar{\ensuremath{\smash{\overline{\mathrm{DR}}}}}

\newcommand\matr[1]{\mathbf{#1}}

\newcommand\cL{{\cal L}}

\newcommand\SW{s_\mathrm{w}}
\newcommand\CW{c_\mathrm{w}}
\newcommand\sw{\SW}
\newcommand\cw{\CW}
\newcommand\MW{M_W}
\newcommand\MZ{M_Z}

\newcommand\dZm[1]{\delta\matr{Z}_{#1}}

\newcommand\ino[1]{\tilde\chi_{#1}}

\newcommand\chapm[1]{\ino{#1}^\pm}

\newcommand\chap[1]{\ino{#1}^+}
\newcommand\cham[1]{\ino{#1}^-}
\newcommand\cha{\chapm}
\newcommand\mcha[1]{m_{\chapm{#1}}}

\newcommand\neu[1]{\ino{#1}^0}
\newcommand\mneu[1]{m_{\neu{#1}}}

\newcommand\refeq[1]{Eq.~(\ref{#1})}
\newcommand\refeqs[1]{Eqs.~(\ref{#1})}

\newcommand\refse[1]{Sect.~\ref{#1}}

\newcommand\citere[1]{Ref.~\cite{#1}}

\newcommand{\gev}{\,\, \mathrm{GeV}}

\newcommand\edz{\tfrac{1}{2}}

\newcommand\FA{\texttt{FeynArts}}
\newcommand\FC{\texttt{FormCalc}}
\newcommand\LT{\texttt{LoopTools}}

\newcommand{\ccn}[1]{\ensuremath{\mbox{CCN}_{#1}}}
\newcommand{\cnn}[3]{\ensuremath{\mbox{CNN}_{#1 #2 #3}}}
\newcommand{\nnn}[3]{\ensuremath{\mbox{NNN}_{#1 #2 #3}}}
\newcommand{\rs}[1]{\ensuremath{\mbox{RS}_{#1}}}

\def\order#1{\ensuremath{{\cal O}(#1)}}
\def\reffi#1{\mbox{Fig.~\ref{#1}}}

\newcommand{\non}{\nonumber}

\def\Ga{\Gamma}

\def\de{\delta}

\newcommand{\MOne}{\ensuremath{M_1}}
\newcommand{\MTwo}{\ensuremath{M_2}}

\definecolor{Orange}{named}{orange}
\definecolor{Purple}{named}{purple}
\definecolor{Lightblue}{cmyk}{0.9,0.1,0.1,0.3}
\definecolor{dgelborange}{cmyk}{0.,0.3,0.5, 0.}
\definecolor{Lila}{rgb}{0.5,0.,1}

\allowdisplaybreaks

\title{Automated choice of the best renormalization scheme}

\ShortTitle{Automated RS choice}

\author{\speaker{S.~Heinemeyer}\\ 
Instituto de F\'isica Te\'orica (UAM/CSIC), 
Universidad Aut\'onoma de Madrid, Cantoblanco, 28049, Madrid, Spain\\
        E-mail: \email{Sven.Heinemeyer@cern.ch}}

\author{F.~v.d.\,Pahlen\\
  Instituto de F\'isica, Universidad de Antioquia, Calle 70 No. 52-21,
  Medell\'in, Colombia\\
E-mail: \email{federico.vonderpahlen@udea.edu.co}}

\abstract{
High-precision predictions in BSM models require calculations at the
loop-level and thus a renormalization of (some of) the BSM parameter.
Here many choices for the renormalization scheme (RS) are possible.
A given RS can be well suited
to yield ``stable'' and ``well behaved'' higher-order corrections in one
part of the BSM parameter space, but can fail completely in other
parts. The latter may not even be noticed numerically if an isolated
parameter point is investigated. 
Here we review a new method for choosing a ``well behaved'' RS.
We demonstrate the
feasibility of our new method in the chargino/neutralino sector of the
Minimal Supersymmetric Standard Model (MSSM), but stress the general
applicability of our method to all types of BSM models.
}

\FullConference{Loops and Legs in Quantum Field Theory (LL2022)\\
                25 - 30 April 2022 \\
                Kloster Ettal, Germany}

\begin{document}


\section{Introduction}
\label{sec:intro}

A reliable investigation of a model beyond the Standard Model (BSM)
requires the inclusion of
higher-order corrections to, e.g., the production cross sections of BSM
particles at the HL-LHC. This in turn requires the renormalization of
the BSM model.
The renormalization of BSM models is much less explored than the
renormalization of the SM. Examples for ``full one-loop
renormalizations'' can be found for the Minimal Supersymmetric Standard Model
(MSSM)~\cite{Stop2decay,MSSMCT},
and the Next-to-MSSM (NMSSM)~\cite{NMSSMCT}. These
analyses showed that many different choices of renormalization schemes
(RS) are possible. This can concern the choice
of the set of to-be-renormalized parameters out of a larger set of BSM
parameters, but can also concern the type of renormalization condition
that is chosen for a specific parameter. 

BSM models naturally possess several new BSM parameters. The number of
new parameters can vary from \order{1} to \order{10}, or even
higher. Often multi-dimensional parameter scans are employed, or methods
such as Markow-Chain Monte-Carlo (MCMC) analyses
to find the phenomenological best-appealing parameters in the
multi-dimensional BSM parameter space. The above mentioned BSM analyses
also demonstrated that a given RS can be well suited
to yield ``stable'' and ``well behaved'' higher-order corrections (more
details will be given below) in one
part of the BSM parameter space, but can fail completely in other
parts. The latter may not even be noticed numerically if only isolated
parameter points are investigated, which is natural in a scan, or MCMC
analyses.
Consequently, the exploration of BSM
models requires a choice of 
a good RS {\em before} the calculation is performed.

An RS ``fails'' if one of the counterterms (or a linear combination of
counterterms) does not (or only marginally) 
depend on the parameter itself, but is rather determined via other parameters
of the model.
This failure can manifest itsel in 
{\it (i)} ``unnaturally'' large higher-order corrections, 
{\it (ii)} large (numerical) differences between \DRbar\ and OS masses, 
{\it (iii)} (numerical) differences between \DRbar\ and OS parameters.
In this work we review a new method how such a situation can be
avoided, i.e.\ how a ``good'' RS can be chosen. This method is based on
the properties of the transformation matrix that connects the various
counter terms with the underlying parameter. This allows a
point-by-point test of all ``available'' or ``possible'' RS, and the
``best'' one can be chosen to perform the calculation.
Our idea is designed to work in all cases of RS choices (in BSM
models).

The numerical examples will be performed within the MSSM, concretely in
the sector of charginos and neutralinos, the supersymmetric (SUSY)
partners of the SM gauge bosons and the 2HDM-like Higgs sector. 
While this constitutes a very specific example, we would like to stress 
the general applicability of our method to all types of BSM models and
types of RS choices.


\section{Renormalization: theoretical considerations and concrete
  implementations}

\subsection{The general idea}
\label{sec:general}

As discussed above, the idea of how to choose a stable and well behaved
RS is generally applicable. However, here we will outline it focusing a
more concrete problem:
in our theory we have $m$ underlying Lagrangian parameters
and $n > m$ particles or particle masses that can be renormalized OS. 
Each choice of $m$ particles renormalized OS defines an
\rs{l}, of which we have $N$ in total. How can one choose the ``best'' \rs{L}?
Our starting point will be the following:
  The masses of the BSM particles under investigation have not (yet)
  been measured. Then we start with $\DRbar$ parameters.
The general idea for the automated choice of the \rs{L} in the
\DRbar\ case can be outlined for two possible levels of refinement. The
first one is called ``semi-OS scheme'', and the second one ``full-OS
scheme'' (where in our numerical examples we will focus on the
latter). The two cases are defined as follows.


\subsubsection*{Semi-OS scheme:}
\begin{enumerate}
\item
  We start with $m$ \DRbar\ parameters, $P_i^{\DRbar}$, from the
  Lagrangian and $N$ \rs{l}.
\item
  For each \rs{l}, i.e. each different choice of $m$ particles
  renormalized OS, we evaluate the corresponding OS parameters
  \begin{align}
    P_{i,l}^{\Os} = P_i^{\DRbar} - \de P_{i,l|{\rm fin}}^{\Os}
    \label{Pilos}
  \end{align}
  with the transformation matrix $\matr{A}^{\DRbar}_l$ (more details
  will be given below).
\item
  It will be argued that a ``bad'' scheme \rs{l}\ has a small or even
  vanishing $|\det\matr{A}^{\DRbar}_l|$.
\item
  Comparing the various $|\det\matr{A}^{\DRbar}_l|$ yields \rs{L}.
\item
  Inserting $P_{i,L}^{\Os}$ into the Lagrangian yields $n$ particle
  masses out of which $m$ are by definition given as their OS values.
  The remaining OS masses have to be determined calculating $n-m$ finite
  shifts.
\item
  The counterterms for the $P_{i,L}^{\Os}$ are already known from
  \refeq{Pilos} as $\de P_{i,L}^{\Os}$ and can be inserted as
  counterterms in a loop calculation.
\end{enumerate}

\noindent
This procedure yields all ingredients for an OS scheme. However, the
OS counterterms $\de P_{i,L}^{\Os}$ and thus also the OS parameters
themselves, $P_{i,L}^{\Os}$, are calculated in terms of
\DRbar\ parameters, i.e.\ one has $\de P_{i,L}^{\Os}(P_i^{\DRbar})$ and
$P_{i,L}^{\Os}(P_i^{\DRbar})$.
This is unsatisfactory for a ``true'' OS scheme, i.e.\ one would like to have
$\de P_{i,L}^{\OS}(P_{i,L}^{\OS})$.
Furthermore, when a \rs{l}\ ``starts to turn bad'' as a function of a
\DRbar\ parameter, large differnces between the $P_{i,l}^{\Os}$ and
$P_i^{\DRbar}$ occur, shedding doubt on the above outlined procedure.
These problems can be circumvented by extending the above scheme to an
evaluation of the counterterms in terms of OS parameters. The general
idea starts as above, but deviates from step~4 on.


\subsubsection*{Full-OS scheme:}
The first two steps are as in the semi-OS scheme. We then continue with
\begin{itemize}
\item[3.]
  Inserting $P_{i,l}^{\Os}$ into the Lagrangian yields $n$ particle
  masses out of which $m$ are by definition given as their $\Os_l$ values.
  The remaining $\Os_l$ masses have to be determined calculating $n-m$ finite
  shifts.
\item[4.]
  \rs{l} is applied again on the OS$_l$ Lagrangian.
\item[5.]
  This yields now OS counterterms in terms of $\Os_l$ parameters,
  \begin{align}
    \de P_{i,l}^{\OS}(P_{i,l}^{\Os})
    \label{PilOS}
  \end{align}
  with the transformation matrix $\matr{A}_l^{\OS}$ (more details will
  be given below).
\item[6.]
  It will be argued that a ``bad'' scheme \rs{l}\ has a small/
vanishing $|\det\matr{A}^{\DRbar}_l|$ and/or $|\det\matr{A}^{\OS}_l|$.
\item[7.]
  Comparing the various
    $  \min \LV |\det\matr{A}^{\DRbar}_l|, |\det\matr{A}^{\OS}_l| \RV $
  yields \rs{L}.
\item[8.]
  The counterterms for the $P_{i,L}^{\OS}$ are already known from
  \refeq{PilOS} as $\de P_{i,L}^{\OS}$ and can be inserted as
  counterterms in a loop calculation.
\end{itemize}
Steps 3-5 could be iterated until convergence is reached. We will not do
this. 


\subsection{Application to the chargino/neutralino sector of the MSSM}

The concrete implementation concerns the calculation of physics
processes with (external) charginos and/or 
neutralinos, $\cha{c} (c = 1,2)$ and $\neu{n} (n = 1, 2, 3, 4)$ at the loop
level. This requieres the choice of a (numerically well behaved) RS.
The possible scheme choices are ($n'' > n' > n$)
\begin{align}
\ccn{n}, \quad \cnn{c}{n}{n'}, \quad \nnn{n}{n'}{n''} 
\quad c = 1,2; \; n,n',n'' = 1,2,3,4~.
\end{align}
Here \ccn{n} denotes a scheme where the two charginos and the
neutralino~$n$, $\neu{n}$, are renormalized OS. \cnn{c}{n}{n'} denotes a
scheme were chargino~$c$, $\cha{c}$, as well as neutralinos~$n,n'$,
$\neu{n}, \neu{n'}$, are renormalized OS. Finally \nnn{n}{n'}{n''} denotes
a scheme with three neutralinos renormalized OS.
For sake of simplicity, in the following we neglect the \nnn{n}{n'}{n''}
schemes. 

To fix our notation we briefly describe the chargino/neutralino sector
of the MSSM. The bilinear term in the Lagrangian is given by, 
\begin{align}
\cL^{\text{bil.}}_{\cham{},\tilde{\chi}^0} &= 
  \overline{\cham{i}}\, \pslash\, \OM \cham{i} 
+ \overline{\cham{i}}\, \pslash\, \OP \cham{i} 
- \overline{\cham{i}}\, [\matr{V}^* \matr{X}^\top \matr{U}^\dagger]_{ij} \,
  \OM \cham{j} 
- \overline{\cham{i}}\, [\matr{U} \matr{X}^* \matr{V}^{\top}]_{ij} \,
  \OP \cham{j} \non \\
&\quad + \frac{1}{2} \LP
  \overline{\neu{k}}\, \pslash\, \OM \neu{k}, 
+ \overline{\neu{k}}\, \pslash\, \OP \neu{k} 
- \overline{\neu{k}}\, [\matr{N}^*\matr{Y} \matr{N}^\dagger]_{kl} \,
  \OM \neu{l} 
- \overline{\neu{k}}\, [\matr{N} \matr{Y}^* \matr{N}^{\top}]_{kl} \,
  \OP \neu{l} \RP~, 
\end{align}
already expressed in terms of the chargino and neutralino mass eigenstates
$\cham{i}$ and $\neu{k}$, respectively, 
and $i,j = 1,2$ and $k,l = 1,2,3,4$.
The mass eigenstates can be determined via unitary 
transformations where the corresponding matrices diagonalize the chargino and
neutralino mass matrix, $\matr{X}$ and $\matr{Y}$, respectively. 

In the chargino case, two $2 \times 2$ matrices $\matr{U}$ and
$\matr{V}$ are necessary for the diagonalization of the chargino mass
matrix~$\matr{X}$, 
\begin{align}
\matr{M}_{\cham{}} = \matr{V}^* \, \matr{X}^\top \, \matr{U}^{\dagger} =
  \begin{pmatrix} m_{\tilde{\chi}^\pm_1} & 0 \\ 
                  0 & m_{\tilde{\chi}^\pm_2} \end{pmatrix}  \quad
\text{with} \quad
  \matr{X} =
  \begin{pmatrix}
    \MTwo & \sqrt{2} \sinb \MW \\
    \sqrt{2} \cosb \MW & \mu
\label{eq:X}
  \end{pmatrix}~,
\end{align}
where $\matr{M}_{\cham{}}$ is the diagonal mass matrix with the chargino
masses $\mcha{1}, \mcha{2}$ as entries, which are determined as the
(real and positive) singular values of $\matr{X}$. 
The singular value decomposition of $\matr{X}$ also yields results for 
$\matr{U}$ and~$\matr{V}$. 

In the neutralino case, as the neutralino mass matrix $\matr{Y}$ is
symmetric, one $4 \times 4$~matrix is sufficient for the diagonalization
\begin{align}
\matr{M}_{\neu{}} = \matr{N}^* \, \matr{Y} \, \matr{N}^{\dagger} =
\text{\bf diag}(m_{\neu{1}}, m_{\neu{2}}, m_{\neu{3}}, m_{\neu{4}})
\end{align}
with
\begin{align}
\matr{Y} &=
  \begin{pmatrix}
    \MOne                  & 0                & -\MZ \, \sw \cosb
    & \MZ \, \sw \sinb \\ 
    0                      & \MTwo            & \quad \MZ \, \cw \cosb
    & -\MZ \, \cw \sinb \\ 
    -\MZ \, \sw \cosb      & \MZ \, \cw \cosb & 0
    & -\mu             \\ 
    \quad \MZ \, \sw \sinb & -\MZ \, \cw \sinb & -\mu              & 0
  \end{pmatrix}~.
\label{eq:Y}
\end{align}
$\MZ$ and $\MW$ are the masses of the $Z$~and $W$~boson, 
$\cw = \MW/\MZ$ and $\sw = \sqrt{1 - \cw^2}$. 
The unitary 4$\times$4 matrix $\matr{N}$ and the physical neutralino
(tree-level) masses $\mneu{k}$ ($k = 1,2,3,4$) result from a numerical Takagi 
factorization of $\matr{Y}$. 

Concerning the renormalization of this sector, the
following replacements of the parameters and the
fields are performed according to the multiplicative renormalization
procedure, which is formally identical for the two set-ups:
\begin{align}
M_1 \; &\to \; M_1 + \de M_1 ~, \quad
M_2 \; \to \; M_2 + \de M_2 ~, \quad
\mu \; \to \; \mu + \de \mu ~, \\
\OMP \chapm{i} \; &\to \; \LB \id + \edz \dZm{\chapm{}}^{L/R} \RB_{ij}
                         \OMP \chapm{j} \qquad (i,j = 1,2)~, \\
\OMP \neu{k} \; &\to \; \LB \id + \edz \dZm{\neu{}}^{/*} \RB_{kl}
                       \OMP \neu{l} \qquad (k,l = 1,2,3,4)~.
\label{dZNeuR}
\end{align}
It should be noted that the parameter counterterms are complex
counterterms which each need two renormalization conditions to be fixed.
The transformation matrices are not renormalized, so that, using the notation 
of replacing a matrix by its renormalized matrix and a counterterm matrix 
\begin{align}
\label{deX}
\matr{X} &\to \matr{X} + \de\matr{X} ~, \quad
\matr{Y} \to \matr{Y} + \de\matr{Y} ~
\end{align}
with
\begin{align}\label{deltaX}
\de\matr{X} &= 
  \begin{pmatrix} \de M_2 & \sqrt{2}\, \de(\MW \Sbe) \\
                  \sqrt{2}\, \de(\MW \Cb) & \de \mu
  \end{pmatrix}~, \\[.4em]
\de\matr{Y} &= 
  \begin{pmatrix} 
      \de M_1 & 0 & -\de(\MZ\sw\Cb) & \de(\MZ\sw\Sbe) \\
      0 & \de M_2 & \de(\MZ\cw\Cb) & -\de(\MZ\cw\Sbe) \\
      -\de(\MZ\sw\Cb) & \de(\MZ\cw\Cb) & 0 & -\de\mu  \\
      \de(\MZ\sw\Sbe) & -\de(\MZ\cw\Sbe) & -\de\mu & 0
  \end{pmatrix}~,
\end{align}
the replacements of the matrices $\matr{M}_{\cham{}}$ and $\matr{M}_{\neu{}}$
can be expressed as
\begin{align}
\matr{M}_{\cham{}} &\to \matr{M}_{\cham{}} + \de\matr{M}_{\cham{}}
   = \matr{M}_{\cham{}} + \matr{V}^* \de\matr{X}^\top \matr{U}^\dagger \\
\label{Mneu}
\matr{M}_{\neu{}} &\to \matr{M}_{\neu{}} + \de\matr{M}_{\neu{}}
   = \matr{M}_{\neu{}} + \matr{N}^* \de\matr{Y} \matr{N}^\dagger~. 
\end{align}
More details on the renormalization can be found in \citere{ccn}.


\subsection{Concrete renormalization in the semi-OS scheme}
\label{sec:semi-os-ren}

We start with \DRbar\ mass matrices for charginos and neutralinos, collectively
denoted as $\matr{X}^{\DRbar}(P_i^{\DRbar})$, depending on the three input
parameters, 
\begin{align}
P_i^{\DRbar} &= M_1^{\DRbar}, M_2^{\DRbar}, \mu^{\DRbar} = \LV p_i^{\DRbar} \RV~.
\end{align}
The mass matrices can be diagonalized, 
\begin{align}
\matr{X}^{\DRbar} \to \matr{M}^{\DRbar} := 
(\matr{N}^{\DRbar})^\dagger \matr{X}^{\DRbar} \matr{N}^{\DRbar}~,
\end{align}
containing on the diagonal two charginos and four neutralino masses, $m_j$.\\
The $\matr{X}^{\DRbar}$ can be renormalized, 
\begin{align}
  \matr{X}^{\DRbar} &\to \matr{X}^{\DRbar} + \de\matr{X}^{\DRbar}(\de P_i^{\DRbar})~\\
\matr{M}^{\DRbar} &\to \matr{M}^{\DRbar} + \de\matr{M}^{\DRbar}(\de P_i^{\DRbar})
  = \matr{M}^{\DRbar} + (\matr{N}^{\DRbar})^\dagger 
                        \de\matr{X}^{\DRbar}(\de P_i^{\DRbar}) \matr{N}^{\DRbar}~.
                        \label{eq.XDRren}
\end{align}
So far, the $\de P_i^{\DRbar}$ are unkown.
The self-energies of the charginos and neutralinos can be written down as
$\SE{j}(P_i^{\DRbar}, \matr{X}^{\DRbar})$.
Now the RS is chosen: \ccn{c}\ or \cnn{c}{n}{n'}.
For each of these $N = 28$ schemes we perform the following. The scheme
is denotes as \rs{l} ($l = 1 \ldots 28$). 
Three renormalized self-energies are chosen to be zero, 
\begin{align}
\rSE{k,l}(P_i^{\DRbar}, \matr{X}^{\DRbar}) = 0~(k = 1,2,3)~,
\end{align}
corresponding to three $\Os$ masses, $m_k^\Os$.
The three renormalized
self-energies yield three conditions on $\de\matr{M}^{\DRbar}_k$,
\begin{align}
\de\matr{M}^{\DRbar}_{k,l} &= f^{\DRbar}_{k,l}(m_{k',l}^{\DRbar}, \SE{k'',l}) 
         + F^{\DRbar}_{k,l}(\de\tb, \de\MZ^2, \ldots) \label{eq.deltaMDR}\\[.5em]
\label{ADRbar}
&\downarrow \matr{A}_l^{\DRbar}\\[.5em]
\de P_{i,l}^\Os &= g^{\DRbar}_{i,l}(m_{k',l}^{\DRbar}, \SE{k'',l})
+ G^{\DRbar}_{i,l}(\de\tb, \de\MZ^2, \ldots)~,
\label{dePilOs}
\end{align}
yielding the $\Os$ values
\begin{align}
P_i^{\DRbar} &\to P_i^{\DRbar} - \de P_{i,l|{\rm fin}}^\Os \; = \; P_{i,l}^\Os~.
\label{PiOs}
\end{align}
It is worth noticing that in the r.h.s.\ of Eq.~(\ref{eq.deltaMDR})
$f_{k,l}$ is linear in $\de P_{i,l}^\Os$, while $F_{k,l}$ only depends on the
counterterm of the remaining model parameters. 
These relations define $\matr{A}_l^{\DRbar}$, the transformation matrix
from the set of mass counterterms to parameter counterterms, 
\begin{align}
\de P_{i,l}^\Os &= (\matr{A}_l^{\DRbar})^{-1}_{ik} 
        \LP \de \matr{M}^{\DRbar}_{k,l}
         - F_{k,l}(\de\tb, \de\MZ^2, \ldots) \RP~.
\label{eq.deltaPOsDR}
\end{align}
$\Os$ masses $m_{k,l}^\Os$ are derived from 
\begin{align}
\matr{X}_l^\Os(P_{i,l}^{\Os}) \to \matr{M}_l^\Os := 
  (\matr{N}_l^\Os)^\dagger \matr{X}_l^\Os(P_{i,l}^{\Os}) \matr{N}^\Os~.
\label{MOs}
\end{align}
The three masses that are not obtained as $\Os$ masses so far can be
evaluated by adding finite shifts to them, see \citere{ccn}.


As discussed above, an RS ``fails'' if one of the counterterms (or a
linear combination of counterterms) does not (or only marginally) 
depend on the parameter itself, but is rather determined via other parameters
of the model.
This is exactly given in our ansatz if the matrix $\matr{A}_l^{\DRbar}$
does not provide a numerically ``well behaved'' transition 
\begin{align}
\de \matr{M}_{k.l}^{\DRbar} \stackrel{\matr{A}_l^{\DRbar}}{\to} \de P^\Os_{i,l}~,
\end{align}
see \refeqs{ADRbar}, suppressing terms involving other
counterterms ($\de\tb$, $\de\MZ^2$, \ldots).
Following the argument of the ``well behaved'' transition, \rs{l} fails if 
$\matr{A}^{\DRbar}_l$ becomes (approximately)
singular, or the normalized determinant, 
\begin{align}
\matr{D}^{\DRbar}_l := \frac{|\det\matr{A}^{\DRbar}_l|}
                           {||\matr{A}^{\DRbar}_l||} \ll 1~,
\end{align}
Conversely, the ``best'' scheme \rs{L} can be chosen via the condition
of the maximum normalized determinant, 
\begin{align}
  \rs{L}^\Os \quad \Leftrightarrow \quad \matr{D}_L^{\DRbar} =
  \max_l \LV \matr{D}_l^{\DRbar} \RV~.
\end{align}


\subsection{Concrete implementation in the full OS renormalization}
\label{sec:full-os-ren}

For each \rs{l} as evaluated in \refse{sec:semi-os-ren}
we now have $\Os$ mass matrices for charginos and neutralinos, collectively
denoted as $\matr{X}^\Os(P_{i,l}^{\Os})$ following \refeq{MOs}. We also
have $\Os$ parameters $P_{i,l}^\Os(P_i^{\DRbar})$ following \refeq{PiOs}
and $\de P_{i,l}^\Os(P_i^{\DRbar})$ following \refeq{dePilOs}.
This is unsatisfactory for a ``true'' OS scheme, i.e.\ one would like to have
$\de P_{i,l}^{\OS}(P_{i,l}^{\OS})$.
Furthermore, when a \rs{l}\ ``starts to turn bad'' as a function of a
\DRbar\ parameter, large differnces between the $P_{i,l}^{\Os}$ and
$P_i^{\DRbar}$ occur, shedding doubt on the above outlined procedure.
These problems can be circumvented by extending the above scheme to an
evaluation of the counterterms in terms of OS parameters.

We start with the $\Os$ parameters obtained in \refse{sec:semi-os-ren},
$P_{i,l}^\Os$. The mass matrices depend on these three input parameters.
Now the renormalization process in \rs{l} is applied again, starting
from the above $\Os$ values. Following the same steps as in
\refse{sec:semi-os-ren}, defining the matrix $\matr{A}_l^\Os$.
As in the case of the semi-OS scheme, a bad \rs{l} is indicated if
in our ansatz if the matrix $\matr{A}_l^\Os$ does not
provide a numerically ``well behaved'' transition 
\begin{align}
\de \matr{M}_{k,l}^\Os \stackrel{\matr{A}_l^\Os}{\to} \de P_{i,l}^\OS~,
\end{align}
and suppressing terms involving other
counterterms ($\de\tb$, $\de\MZ^2$, \ldots).
Following the argument of the ``well behaved'' transition, \rs{l} fails if 
$\matr{A}^{\DRbar}_l$ or $\matr{A}^\Os_l$ become (approximately)
singular, or the normalized determinant, 
\begin{align}
\label{eq:matrA}
\matr{D}^{\DRbar}_l := \frac{|\det\matr{A}^{\DRbar}_l|}
                                 {||\matr{A}^{\DRbar}_l||} \ll 1
\quad\mbox{or}\quad
\matr{D}^{\Os}_l := \frac{|\det\matr{A}^{\Os}_l|}
                                 {||\matr{A}^{\Os}_l||} \ll 1~,
\end{align}
equivalent to
$  \matr{D}_l^\OS := 
  \min\LV \matr{D}^{\DRbar}_l, \matr{D}^{\Os}_l \RV \ll 1$~.
Conversely, the ``best'' scheme \rs{L} can be chosen via the condition
of the maximum normalized determinant, 
\begin{align}
  \rs{L}^\OS \quad \Leftrightarrow
      \quad \matr{D}^\OS_L = \max_l \LV \matr{D}^\OS_l \RV~.
\end{align}
\noindent
Now all ingrediences for physics calculations are at hand.
{\it (i)}  The physical parameters $P_{i,L}^{\OS}$ are given via the OS analogon
  to \refeq{PiOs}.
{\it (ii)}  The counterterms for the $P_{i,L}^{\OS}$ are known from the OS analogon to
\refeq{dePilOs} as $\de P_{i,L}^{\OS}$ and can be inserted as
counterterms in a loop calculation.
{\it (iii)}  Inserting $P_{i,L}^{\OS}$ into the Lagrangian yields six particle
masses out of which three are by definition given as their $\OS$ values.
The remaining $\OS$ masses have to be determined calculating three finite
shifts, see \citere{ccn}.


\section{Numerical example}

\begin{figure}[h!]
\begin{center}
  \includegraphics[width=0.53\textwidth]{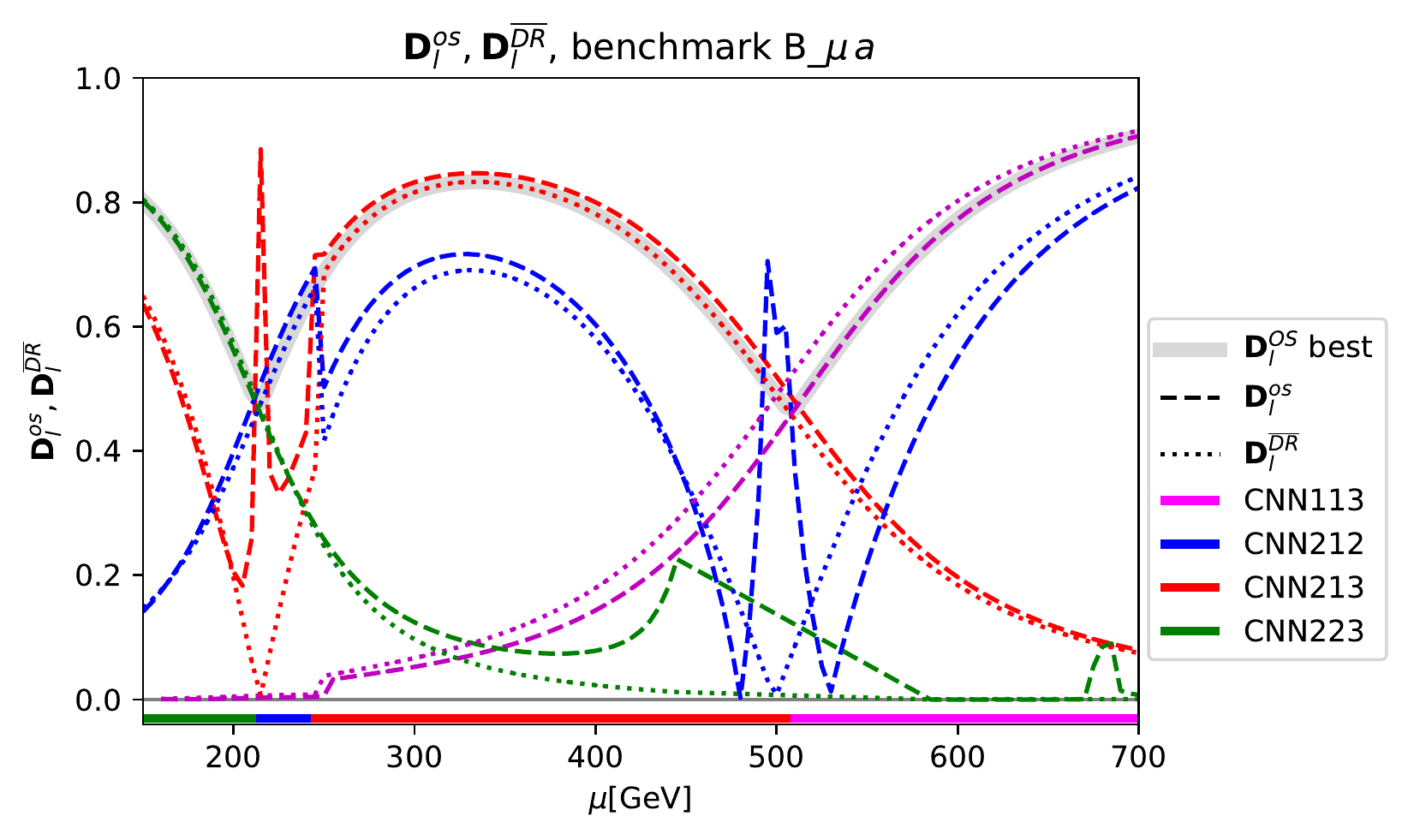}\\
  \includegraphics[width=0.43\textwidth]{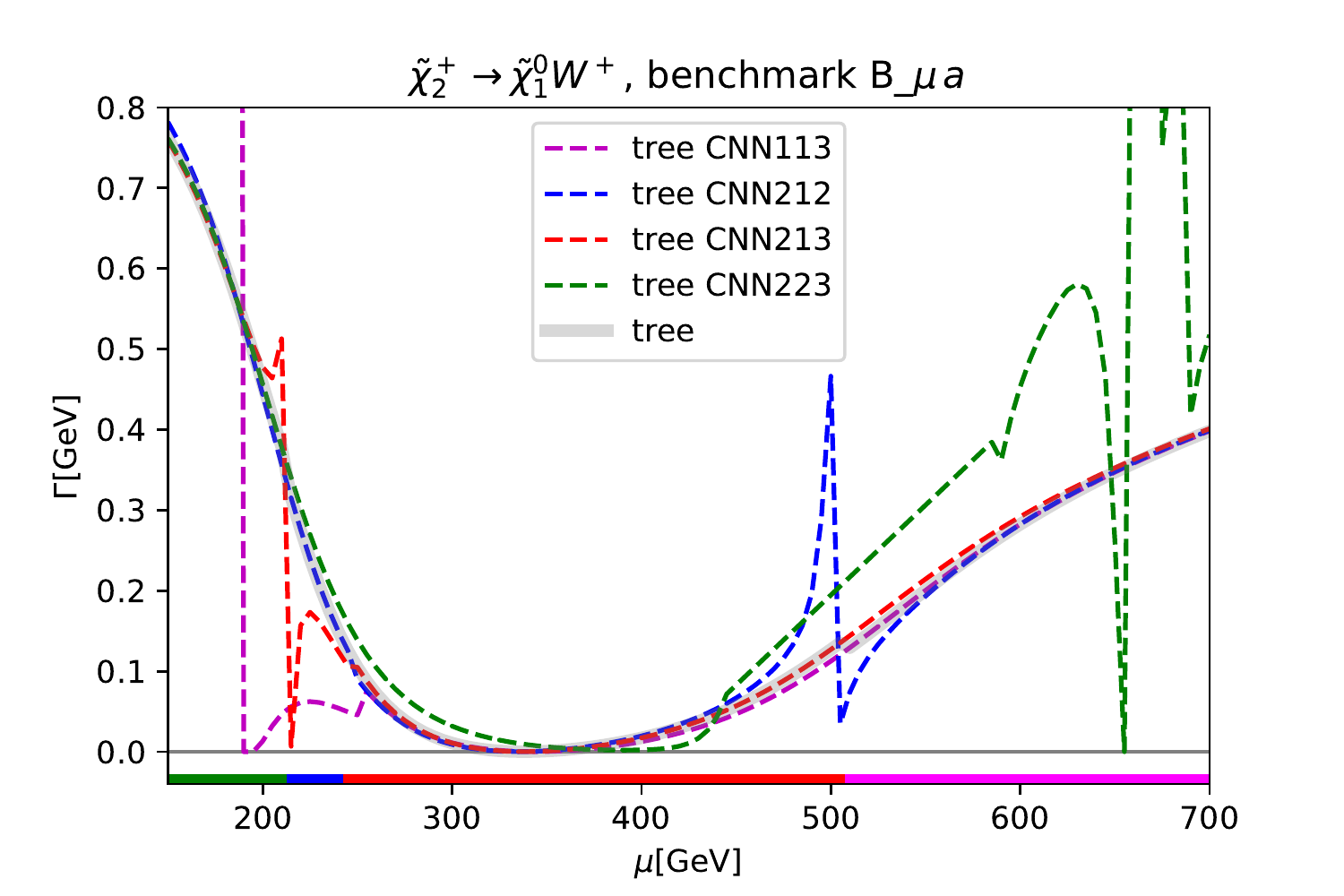}
  \includegraphics[width=0.43\textwidth]{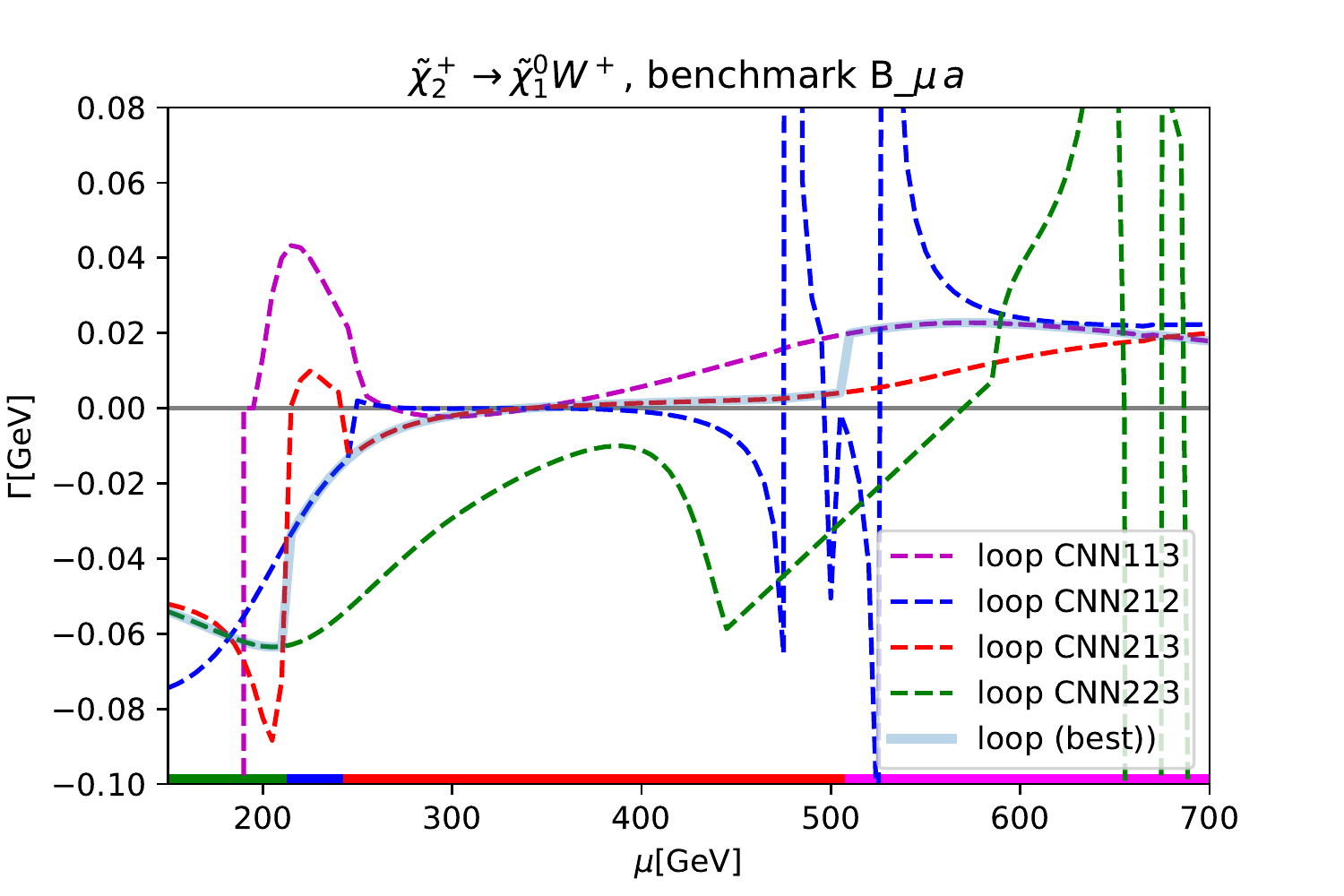}
  \includegraphics[width=0.43\textwidth]{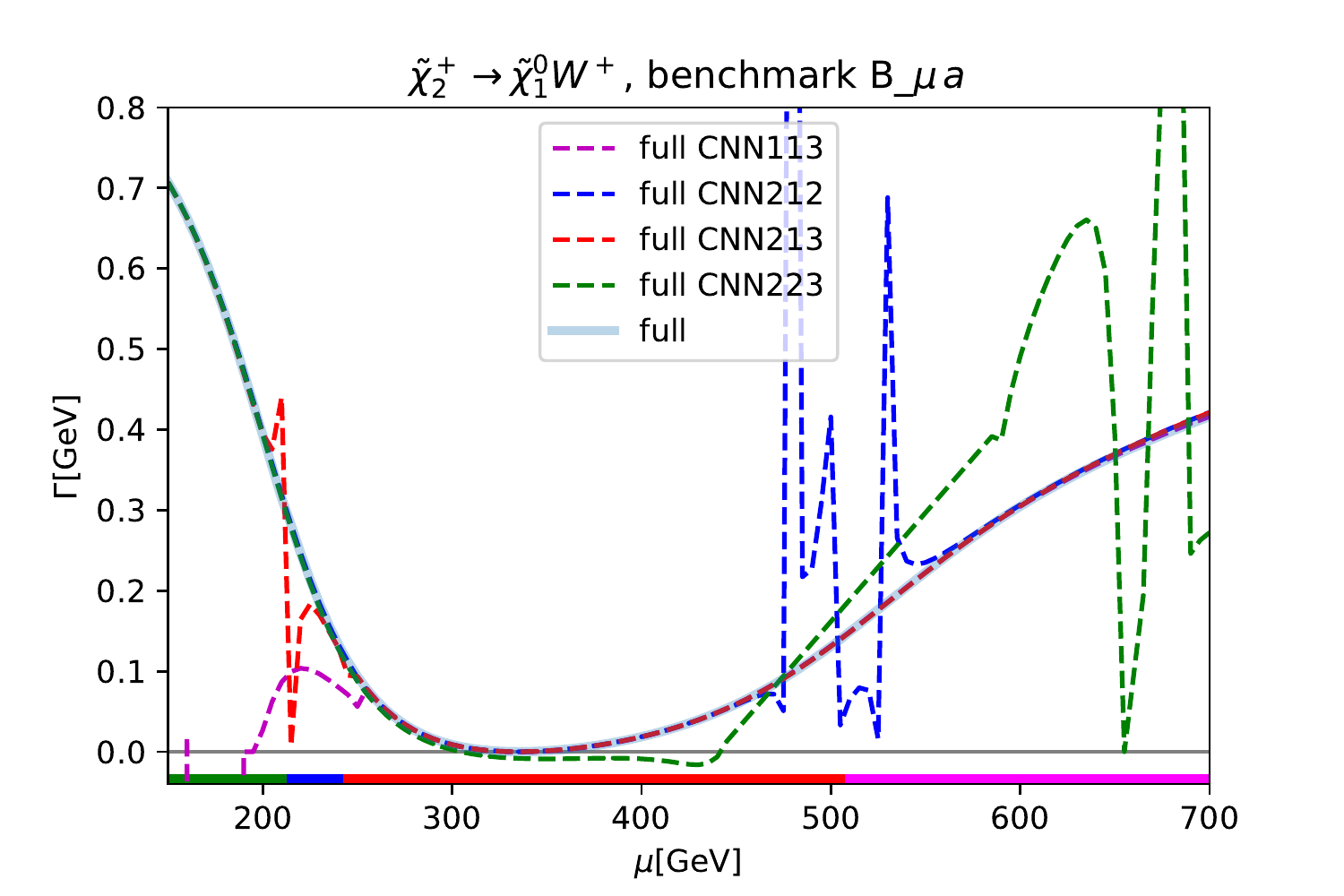}
  \includegraphics[width=0.43\textwidth]{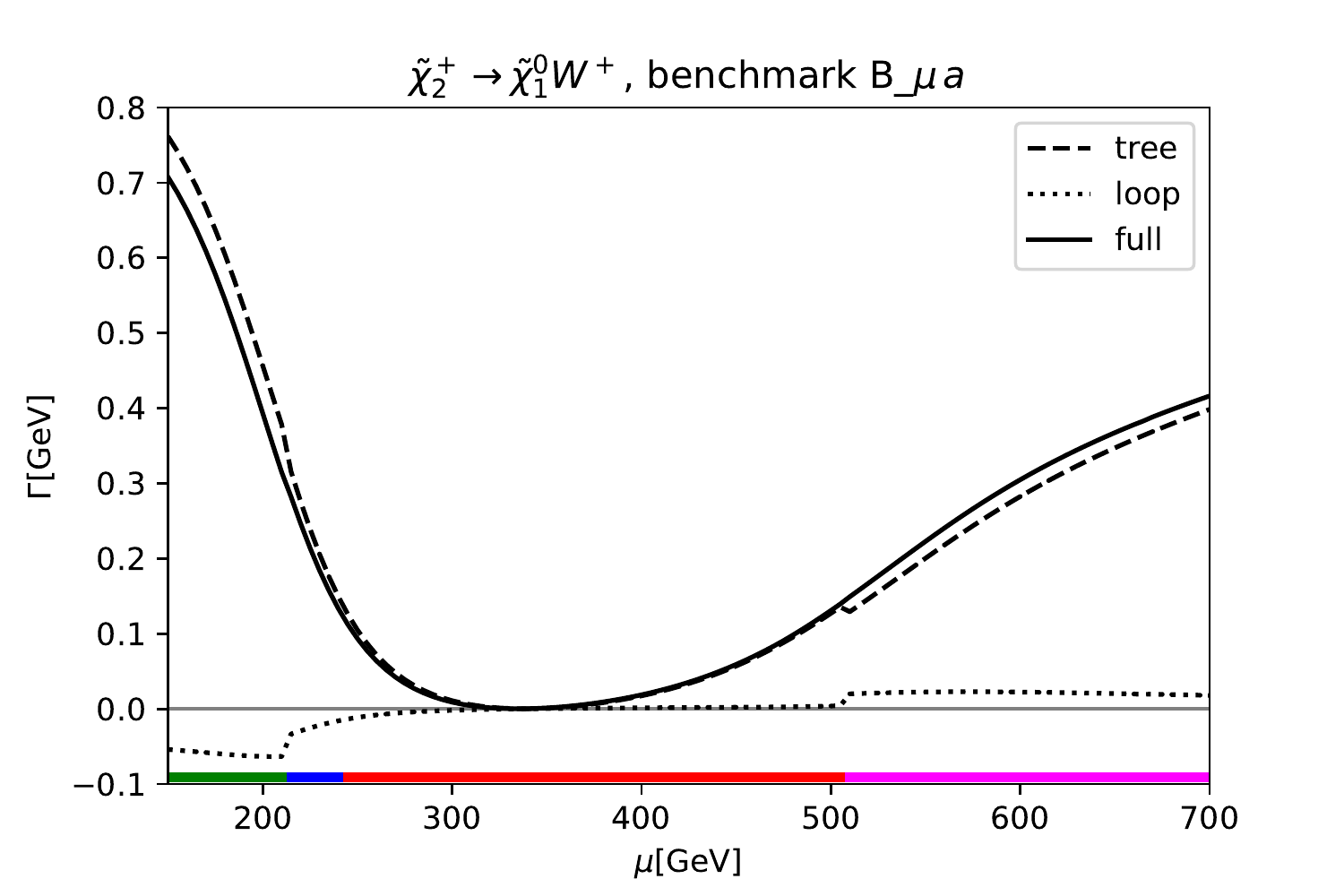}
  \caption{$\Ga(\chap{2} \to \neu{1} W^+)$ as a function of $\mu$
 for $M_1 = 200 \gev$, $M_2 = 500 \gev$ and $\tb = 10$ (see text).
}
\label{fig:2022_C2N1W_mu_test}
\end{center}
\vspace{-2em}
\end{figure}

As numerical example of the application of our procedure we show in
\reffi{fig:2022_C2N1W_mu_test} the results for the
decay width for $\Ga(\chap{2} \to \neu{1} W^+)$ as a function of $\mu$
for $M_1 = 200 \gev$, $M_2 = 500 \gev$ and $\tb = 10$.
The results were obtained using the \FA/\FC/\LT\
set-up~\cite{feynarts2,feynarts3,formcalc1} with the 
MSSM model file as defined in \citere{MSSMCT}. 
The upper plot shows  the normalized determinants
$\matr{D}^{\DRbar}_l $ (dotted) and $\matr{D}^{\Os}_l $ (dashed),
see \refeq{eq:matrA} in four colors for the four ``best RS''.
The results of the ``selected best RS'' are overlaid with a gray band.
The horizontal colored bar indicates this best RS for the corresponding
value of  $\mu$,  
following  the same color coding as the curves:
CNN$_{223}$ for $\mu \lsim 210\gev$, 
CNN$_{212}$ for $215 \gev \lsim \mu \lsim 240\gev$, 
CNN$_{213}$ for $245 \gev \lsim \mu \lsim 505\gev$, 
CNN$_{113}$ for $510 \gev \lsim \mu$. 
In this example the selected best scheme has determinants larger than
$\sim 0.5$, indicating that the counter terms can be determined reliably.
The middle left figure shows the tree results for the same four selected
RS as colored
dashed lines, and the results of the ``selected best RS'' are again overlaid
with a gray band.
One can observe that where a scheme is chosen, the tree level width
behaves ``well'' and smooth. It reaches zero at $\mu \sim 330 \gev$
because the involved tree-level coupling has an (accidental) zero
crossing. On the other hand, outside the selected interval the
tree-level result behave highly irregular, induced by the shifts in the
mass matrices to obtain OS masses.
The middle right plot shows the ``loop plus real photon emission''
results with the same color coding as in the middle left plot.
As for the tree-level result one sees that where a scheme is chosen the
loop corrections behave smooth and the overall size stays at the level
of $\sim 10\%$ or less compared to the tree-level result. As above,
outside the chosen interval the loop corrections take irregular values,
which sometimes even diverge, owing to a vanishing determinant.
The lower left plot, using again the same color coding, shows the sum of
tree and higher-order corrections, i.e.\ of the two previous plots. The
same pattern of numerical behavior can be observed. The chosen scheme
yields a reliable higher-order corrected result, whereas other schemes
result in highly irregular and clearly unreliable results. This is
summarized in the lower right plot, where show the selected tree-level
result as dashed line, the loop result as dotted, and the full result as
solid line. The overall behavior is completely well-behaved and
smooth. A remarkable feature can be observed at $\mu \sim 500
\gev$. Here the selected tree-level result has a kink, because of a
change in the shift in the OS values of the involved chargino/neutralino
masses, caused by the change from switching from \cnn{2}{1}{3} to
\cnn{1}{1}{3}. However, the loop corrections contain also a
corresponding kink, leading to a completely smooth full one-loop result.

This shown example demomstrates the power of the new algorythm used to
select {\it beforehand} the best RS out of many. It also demonstrates
that without such a scheme choice completely unreliable results can be
obtained. 


\mbox{}\vspace{-2em}
\subsubsection*{Acknowledgements}

S.H.\ thanks the  organizers of L\&L\,2022 for the  invitation and the
(as always!) inspiring atmosphere.
The work of S.H.\ has received financial support from the
grant PID2019-110058GB-C21 funded by
MCIN/AEI/10.13039/501100011033 and by "ERDF A way of making Europe".
MEINCOP Spain under contract PID2019-110058GB-C21 and in part by
by the grant IFT Centro de Excelencia Severo Ochoa CEX2020-001007-S
funded by MCIN/AEI/10.13039/501100011033.


\newcommand\jnl[1]{\textit{\frenchspacing #1}}
\newcommand\vol[1]{\textbf{#1}}

\end{document}